\documentclass[conference]{IEEEtran}
\usepackage{cite}
\ifCLASSINFOpdf
\usepackage[pdftex]{graphicx}
\graphicspath{./figures}
\DeclareGraphicsExtensions{.pdf,.jpeg,.png}
\else
\fi
\usepackage[cmex10]{amsmath}
\usepackage{array}
\usepackage{url}
\usepackage[utf8]{inputenc}  
\usepackage{citesort}
\usepackage[T1]{fontenc} 
\usepackage{mathtools}
\usepackage{amsfonts}
\usepackage{amssymb}
\usepackage{graphicx}

\DeclareUnicodeCharacter{00A0}{~}

\def \({\left(}
\def \){\right)}
\def \[{\left[}
\def \]{\right]}
\newcommand{\tbf}[1]{{\textbf{#1}}}

\newcommand{\txt}[1]{\text{#1}}
\newcommand{\defeq}{\vcentcolon=}

\newcommand{\bF}{{\textbf {F}}}

\newcommand{\bx}{{\textbf {x}}}

\newcommand{\by}{{\textbf {y}}}
\newcommand{\bz}{{\textbf {z}}}

\newcommand{\bs}{{\textbf {s}}}

\newcommand{\bxi}{{\boldsymbol{\xi}}}

\newcommand{\be}{\begin{equation}}
\newcommand{\ee}{\end{equation}}
\newcommand{\bea}{\begin{eqnarray}}
\newcommand{\eea}{\end{eqnarray}}

\newtheorem{theorem}{Theorem}[section]
\newtheorem{lemma}[theorem]{\textbf{Lemma}}
\newtheorem{thm}[theorem]{\textbf{Theorem}}

\newtheorem{corollary}[theorem]{\textbf{Corollary}}
\newtheorem{definition}[theorem]{\textbf{Definition}}

\begin{document}
\title{Threshold Saturation of Spatially Coupled Sparse Superposition Codes for All Memoryless Channels}
\author{\IEEEauthorblockN{Jean Barbier, \emph{Member IEEE}, Mohamad Dia and Nicolas Macris, \emph{Member IEEE}.}\\
\IEEEauthorblockA{Laboratoire de Th\'eorie des Communications, Ecole Polytechnique F\'ed\'erale de Lausanne.\\ \{jean.barbier, mohamad.dia, nicolas.macris\}@epfl.ch}}
\maketitle
\IEEEpeerreviewmaketitle
\begin{abstract}
We recently proved threshold saturation for spatially coupled sparse superposition codes on the additive white Gaussian noise channel \cite{barbierDiaMacris_isit2016}. Here we generalize our analysis to a much broader setting. We show for any memoryless channel that spatial coupling allows generalized approximate message-passing (GAMP) decoding to reach the potential (or Bayes optimal) threshold of the code ensemble. Moreover in the large input alphabet size limit: $i)$ the GAMP algorithmic threshold of the underlying (or uncoupled) code ensemble is simply expressed as a Fisher information; $ii)$ the potential threshold tends to Shannon's capacity. Although we focus on coding for sake of coherence with our previous results, the framework and methods are very general and hold for a wide class of generalized estimation problems with random linear mixing.
\end{abstract}
\section{Introduction}
Sparse superposition (SS) codes were developed for reliable communication over the additive white Gaussian noise (AWGN) channel \cite{barron2010sparse} and were proven to be capacity-achieving for this channel when power allocation and iterative decoding are employed \cite{JosephB14,barron2012high}. Later on, the approximate message-passing (AMP) decoder was 
introduced in \cite{barbier2014replica} and spatial coupling (SC) constructions (also combined with efficient Hadamard-based operators) were presented in \cite{barbierSchulkeKrzakala,BarbierK15}. These SC constructions have many similarities with those introduced in the context of compressed sensing \cite{KrzakalaMezard12,CaltagironeZ14}, the first successful application of SC to dense systems. An independent line of work also studying the AMP decoder, but using power allocation instead of SC, is presented in \cite{rush2015capacity}.

It appears that SC-SS codes have much better performances than power allocated ones \cite{BarbierK15}. This motivated the initiation of their rigorous study \cite{barbierDiaMacris_isit2016} using the potential method, originally developed for low density parity check codes \cite{YedlaJian12,PfisterMacrisBMS,6887298}. In \cite{barbierDiaMacris_isit2016} we showed that $i)$ threshold saturation occurs, i.e. minimum mean square error (MMSE) performance is reached using SC and AMP decoding, and $ii)$ the potential threshold (above which AMP decoding is not possible without using SC or power allocation) tends to capacity in the large alphabet size limit, and this even without power allocation. 

These encouraging results (obtained for the AWGN) naturally led us to study a general setting that includes all memoryless channels and any input signal model that factorizes over $B$-dimensional ($B$-d) sections $p_0(\bs)=\prod_{l=1}^L p_0(\bs_l)$, $\bs_l\in \mathbb{R}^B$. 

The present analysis is also based on the potential method. 
The correct potential and associated \emph{state evolution} (SE) for the present setting can be ``guessed'' using the replica method. Alternatively, one can ``integrate'' the SE associated with the GAMP algorithm in the vectorial setting. The GAMP equations were originally derived for scalar estimation \cite{rangan2011generalized}, but their extension to the present vectorial setting is immediate.


%
\section{Code ensembles}\label{sec:codeens}
%
%
%
In the sequel, the shorthands $[a_1 : a_n]$ and $\{a_1 : a_n\}$ refer to $[a_1, \dots, a_n]$ and $\{a_1, \dots, a_n\}$ respectively. 
The probability distribution of a Gaussian random variable $x$ with mean $m$ and variance $\sigma^2$ is denoted $\mathcal{N}(x\vert m, \sigma^2)$.

Let us start defining \textbf{\emph{the underlying ensemble}} of SS codes for transmission over a generic memoryless channel. The \emph{information word} or \emph{message} is a vector made of $L$ \emph{sections}, $\bs = [\bs_1 : \bs_L]$. Each section is a $B$-d vector with a single non-zero component
equal to $1$. $B$ is the \emph{section size} (or alphabet size) and we set $N=LB$. For example if $(B=3,L=4)$, then a valid message could be $\tbf s = [001,010,100,010]$. We consider random linear codes generated by a fixed \emph{coding matrix} $\bF\in \mathbb{R}^{M \times N}$ drawn from the ensemble of random matrices with i.i.d real Gaussian entries
with distribution $\mathcal{N}(\cdot|0, 1/L)$. The \emph{codeword} $\bF\bs\in \mathbb{R}^{M}$ and the cardinality of the code is $B^L$. Hence, the (design) rate is $R=L\log_2(B)/M = N\log_2(B)/(M B)$. The code is thus specified by $(M, R, B)$. The rate $R$ can be linked to the ``measurement rate'' $\alpha$, used in the compressive sensing literature \cite{KrzakalaMezard12}, by $\alpha \defeq M/N = \log_2(B)/(BR)$. 

We want to communicate through a known memoryless channel $W$. This requires to map the codeword components $[\bF\bs]_\mu\in \mathbb{R}$ onto the input alphabet of $W$. Call $\pi$ this map (see Sec.~\ref{sec:larg_B} for various examples). The concatenation of $\pi$ and $W$ can be seen as an \emph{effective memoryless channel} $P_{\text{out}}$, such that $P_{\text{out}}(y_\mu|[\bF\bs]_\mu)\defeq W(y_\mu|\pi([\bF\bs]_\mu))$. In the present framework, it is more convenient to work with this effective memoryless channel $P_{\text{out}}(\by|\bF\bs) = \prod_{\mu = 1}^M P_{\text{out}}(y_\mu|[\bF\bs]_\mu)$, from which the receiver obtains the noisy channel observation $\by$.

%
%
%
%

We now present \textbf{\emph{the spatially coupled ensemble}} of SS codes. We consider SC codes based on coding 
matrices in $\mathbb{R}^{M\times N}$ made of $\Gamma\times \Gamma$ blocks indexed by $(r,c)$, each with $N/\Gamma$ columns and $M/\Gamma =\alpha N/\Gamma$ rows. This ensemble of matrices is parametrized by $(M,R,B,\Gamma,w,g_w)$, where $w$ is the \emph{coupling window} and $g_w$ is the \emph{design function}. 
This is any function verifying $g_w(x) = 0$ if $|x|>1$ and $g_w(x)\ge g_0>0$ else, which 
is Lipschitz continuous on its support with Lipschitz constant $g_*$ independent of $w$. From $g_w$, we construct the \emph{variances} of the blocks: the i.i.d entries inside the block $(r,c)$ are distributed as $\mathcal{N}(0,J_{r,c}/L)$, where $J_{r,c} \defeq \gamma_r \Gamma  g_w((r-c)/w)/(2w+1)$. Here $\gamma_r$ enforces
the ariance normalization $\sum_{c=1}^{\Gamma}J_{r,c}/\Gamma = 1 \ \forall \ r$. This normalization induces homogeneous power over the codeword components, i.e. $[\bF\bs]_\mu^2 \to 1 \ \forall \ \mu$ as $L\to\infty$. The detailed SC construction is explained in \cite{barbierDiaMacris_isit2016}. 

The SC matrix structure naturally induces a block structure in the message, $\bs=[\bs_1 : \bs_\Gamma]$. In each of these blocks there are $L/\Gamma$ sections. We assume that the sections in the first and last $4w$ blocks of the message are known by the decoder. 
This $\emph{seed}$ initiates a decoding wave in the SC code that propagates inward through the entire message. The seed induces a rate loss in the \emph{effective rate} $R_{\text{eff}} = R (1-\frac{8w}{\Gamma})$ of the code, but this loss vanishes as $\Gamma \rightarrow \infty$.
\section{State evolution and potential formulation}\label{sec:stateandpot}
The decoder is the GAMP algorithm, a generalization of AMP to generic memoryless channels, introduced for estimation of scalar signals with i.i.d components \cite{rangan2011generalized}. In the present context the message components are correlated through $p_0(\bs_l)$, therefore we extend GAMP to cover this vectorial setting (similarly to \cite{BarbierK15} for AMP).
We first give the SE equations associated with the underlying and SC ensembles. SE is conjectured to track the performance of the vectorial extension of the GAMP decoder (see Sec.~\ref{sec:openChallenges}). We then define an appropriate {\it potential function} for each ensemble. 
\subsection{State evolution}
The goal is to iteratively compute the average mean square error (MSE) $\tilde E^{(t)} \defeq \mathbb{E}_{\bs, \by}[\frac{1}{L}\sum_{l=1}^L \|\hat{\bs}_l^{(t)} - \bs_l\|_2^2]$ of the GAMP estimate $\hat{\bs}^{(t)}$ at iteration $t$. We first need some definitions.
\begin{definition}[Effective noise] \label{def:effNoise}
Let us define the effective noise variance $\Sigma(E)^{2}$ by the relation
\begin{align}
\Sigma(E)^{-2} \defeq \frac{\mathbb{E}_{p\vert E} [\mathcal{F}(p|E)]}{R},
\end{align}
where the expectation $\mathbb{E}_{p\vert E}$ is w.r.t $\mathcal{N}(p|0,1-E)$ and 
\begin{align}
\mathcal{F}(p|E) \defeq \int dy f(y|p,E) (\partial_x \ln f(y|x,E))^2_{x=p}
\end{align}
is the Fisher information of $p$ associated with the distribution 
$f(y|p,E) \defeq \int du P_{\text{out} }(y|u) \mathcal{N}(u|p,E)$.
\end{definition}
\begin{lemma} \label{lemma:SigmaIncreases}
$\Sigma(E)^2$ is non negative and increasing with $E$.
\end{lemma}
\begin{IEEEproof}
Positivity of the Fisher information implies $\Sigma(E)^2 \ge 0$. The proof that it is increasing is a straightforward application of the data processing inequality for Fisher information (Corollary 6 in \cite{fisherInfoProperties}).
\end{IEEEproof}

From now on, $\bs\sim p_0(\bs)$ and $\bz\sim \mathcal{N}(\bz|0,\tbf{I})$ are $B$-d random vectors and $z\sim\mathcal{N}(z|0,1)$, with expectations noted $\mathbb{E}_{\bs,\bz}$, $\mathbb{E}_{z}$.
\begin{definition}[Denoiser]
The denoiser $g_{\text{in},i}(\bs,\bz,\Sigma)$ is the MMSE estimator of the $i$-th component of a section $\bs$ sent through an effective AWGN channel with a noise $\mathcal{N}(\bxi|0, \tbf{I}\,\Sigma^2/\log_2(B))$. Note that the effective AWGN channel is induced by the code construction and depends on the effective channel $P_{\text{out}}$ only through $\Sigma$. For any $B$-d prior, we have for $i \in \{1:B\}$
\begin{align}
g_{\text{in},i}(\bs,\bz,\Sigma) \defeq \frac{\int d{\bx}\, p_0(\bx) \theta(\bx,\bs,\bz,\Sigma) x_i}{ \int d{\bx} \,p_0(\bx) \theta(\bx,\bs,\bz,\Sigma) },
\end{align}
where $\theta(\bx,\bs,\bz,\Sigma) \defeq \exp\big(-\frac{\|\bx - (\bs +\bz \Sigma/\sqrt{\log_2(B)})\|_2^2}{2\Sigma^2/\log_2(B)} \big)$. Using the prior $p_0(\bx)=\frac{1}{B}\sum_{i=1}^B \delta_{x_i,1}\prod_{j\neq i}^{B-1} \delta_{x_j, 0}$, one recovers the denoiser of SS codes \cite{barbierDiaMacris_isit2016}. 
\end{definition}
\begin{definition}[SE of the underlying system] \label{def:SE}
The SE operator of the underlying system is the average MSE associated with the MMSE estimator of the effective channel, 
\begin{align}
T_{\rm u}(E) \defeq \mathbb{E}_{\bs, \bz}\Big[\sum_{i=1}^B (g_{\text{in},i}(\bs,\bz,\Sigma(E)) - s_i)^2\Big] .
\end{align}
%
The SE tracking the performance of the GAMP decoder is $\tilde E^{(t+1)} = T_{\rm u}(\tilde E^{(t)})$ for $t\geq 0$ and is initialized with $\tilde E^{(0)}=1$.
\end{definition}
%

The existence of a fixed point is ensured by the monotonicity and boundedness of the SE iterations, see Sec.~\ref{sec:proofsketch}.

\begin{definition}[MSE Floor] The MSE floor $E_0$ is the fixed point reached from trivial initial condition, $E_0 = T_{\rm u}^{(\infty)}(0)$. 
\end{definition}
\begin{definition}[Bassin of attraction]
The basin of attraction of the MSE floor $E_0$ is 
$\mathcal{V}_0 \defeq \big\{ E \ \! |\ \! T_{\rm u}^{(\infty )}(E) = E_0 \big\}$.
\end{definition}
\begin{definition} [Threshold of underlying ensemble]
The GAMP threshold is $R_{\rm u} \defeq {\rm sup}\{R>0\ \! |\ \! T_{\rm u}^{(\infty)}(1) = E_0\}$. 
\end{definition}

For the present system, one can show that the only two possible fixed points are $T_{\rm u}^{(\infty)}(0)$ and $T_{\rm u}^{(\infty)}(1)$. For $R<R_{\rm u}$, there is only one fixed point, namely the ``good'' one
$T_{\rm u}^{(\infty)}(0)=E_0$, 
and as the section size $B$ increases $E_0$ and the section error rate (that is the fraction of wrongly decoded sections) vanish. Instead if $R>R_{\rm u}$, the GAMP decoder is blocked by the ``bad'' fixed point $T_{\rm u}^{(\infty)}(1)\neq E_0$.

For a SC system, the performance of GAMP is described by
an average {\it MSE profile} $[\tilde E_c^{(t)} \, \ |\, \ c= 1: \Gamma]$ along the ``spatial dimension'' indexed by the blocks of the message. To reflect the seeding at the boundaries, we enforce the \emph{pinning condition} $\tilde E_c^{(t)}=0$ for $c\in\{1: 4w\}\cup\{\Gamma-4w+1 : \Gamma\}$, at all times. Elsewhere, $\tilde E_c^{(t)} \defeq \mathbb{E}_{\bs, \by}[\frac{\Gamma}{L}\sum_{l\in c} \|\hat{\tbf s}_l^{(t)} - {\tbf s}_l\|_2^2]$, where the sum $l\in c$ is over the set of indices of the $L/\Gamma$ sections composing the $c$-th block of $\bs$. It turns out that the change of variables $E_r^{(t)} \defeq \frac{1}{\Gamma} \sum_{c=1}^{\Gamma} J_{r,c} \tilde E_c^{(t)}$ makes the problem mathematically more tractable. $\tbf{E}$ is called a \emph{profile}. The pinning condition becomes $E_r^{(t)}=0$ for $r\in\mathcal{R}\defeq\{1: 3w\}\cup\{\Gamma-3w+1: \Gamma\}$, and at all times. In order to define the SE of the SC system, we need first the following definition.
\begin{definition}[Per-block effective noise]
The per-block effective noise variance $\Sigma_{c}({\tbf E})^2$ is $\forall \,c \in\{1: \Gamma\}$ defined by
\begin{align}\label{equ:sigmac}
\Sigma_{c}({\tbf E})^{-2} \defeq \sum_{r=1}^{\Gamma} \frac{J_{r,c}}{\Gamma\Sigma(E_r)^2}=\sum_{r=1}^{\Gamma} \frac{J_{r,c}}{R\Gamma} \mathbb{E}_{p\vert E_r} [\mathcal{F}(p|E_r)].
\end{align}
\end{definition}
\begin{definition}[SE of the coupled system] \label{def:SEc}
%
%
The vector valued coupled SE operator is defined componentwise as
%
%
\begin{align}
[T_{\rm c}({\tbf E})]_r \defeq \sum_{c=1}^{\Gamma} \frac{J_{r, c}}{\Gamma}\mathbb{E}_{\bs, \bz}\Big[\sum_{i=1}^B (g_{\text{in},i}(\bs,\bz,\Sigma_{c}({\tbf E} )) - s_i)^2\Big] .
\end{align}
The SE for $r\notin \mathcal{R}$ then reads $E_r^{(t+1)} = [T_{\rm c}({\tbf E}^{(t)})]_r$ for $t\geq 0$.
%
For $r\in \mathcal{R}$, the pinning condition $E_r^{(t)} =0$ is enforced at all times. SE is initialized with $E_r^{(0)}=1$ for $r\notin \mathcal{R}$. 
\end{definition}
Let ${\tbf E}_0 \defeq [E_r=E_0 \ \!|\ \! r=1: \Gamma]$ be the \emph{MSE floor profile}. 
\begin{definition} [Threshold of coupled ensemble]\label{def:AMPcoupled}
The GAMP threshold of the SC system is defined as $R_{\text c} \defeq {\liminf}_{\Gamma, w\to \infty} {\rm sup}\{R>0\ \! |\ \! T_{\text{c}}^{(\infty)}(\boldsymbol{1}) \prec \tbf E_0\}$
where $\boldsymbol{1}$ is the all ones vector. Here the ${\liminf}_{\Gamma, w \to \infty}$ is taken 
along sequences where {\it first} $\Gamma \to \infty$ and {\it then} $w\to\infty$ (see Definition~\ref{def:degradation} for the meaning of $\prec$). 
\end{definition}
\subsection{Potential formulation}\label{subsec:potentials}
The fixed point equations associated with SE can be reformulated as stationary point equations of {\it potential functions} (obtained from the replica method \cite{barbier2014replica} or integrating SE). 
\begin{definition}[Potentials]
The potential of the underlying ensemble is $F_{\rm u}( E) \defeq U_{\rm u}(E) - S_{\rm u}( \Sigma(E))$, with
\begin{align*}
\begin{cases}
U_{\rm u}(E) \defeq -\frac{E}{2\ln(2)\Sigma(E)^2} - \frac{1}{R} \mathbb{E}_z[\int dy\, \phi \log_2(\phi)], \\
S_{\rm u}( \Sigma( {E})) \defeq \mathbb{E}_{\bs,\bz}[\log_B(\int d{\bx} \,p_0(\bx) \theta(\bx,\bs,\bz,\Sigma(E)))],
\end{cases}
\end{align*}
where $\phi(y|z,E) \defeq \int dx P_{\text{out}}(y|x)\mathcal{N}(x|z \sqrt{1 - E}, E)$.
The potential of the SC ensemble is $F_{\text{c}}({\tbf E}) \defeq U_{\text{c}}({\tbf E}) - S_{\text{c}}({\tbf E})$ 
where $U_{\text{c}}(\tbf{E}) \defeq \sum_{r=1}^{\Gamma} U_{\rm u}(E_r)$ and $S_{\text{c}}(\tbf{E}) \defeq\sum_{c=1}^{\Gamma} S_{\rm u}( \Sigma_c({\tbf E}))$.
\end{definition}
\begin{definition}[Free energy gap]
The free energy gap is $\Delta F_{\rm u} \defeq {\rm inf} _{E \notin \mathcal{V}_0 } (F_{\rm u}( E) - F_{\rm u}(E_0))$, with the convention that the infimum over the empty set is $\infty$ (i.e. when $R < R_{\rm u}$).
\end{definition}
\begin{definition}[Potential threshold]
\label{def:potThresh}
The potential threshold is $R_{\rm pot} \defeq {\rm sup}\{R>0\ \! |\ \! \Delta F_{\rm u} > 0\}$. 
\end{definition}

The next Lemma links the potential and SE formulations. 
\begin{lemma}\label{lemma:fixedpointSE_extPot}
One can show that if $T_{\text u}(\mathring E) = \mathring E$, then $\frac{\partial F_{\rm u}}{\partial E}|_{\mathring E} =0$. Similarly for the SC system, if $[T_{\rm c}(\mathring{{\tbf E}})]_r = \mathring{E}_r$ $\forall \ r\in \mathcal{R}^{\text{c}} =\{3w+1 : \Gamma-3w\}$ then $\frac{\partial F_{\text{c}}}{\partial E_r}|_{\mathring{{\tbf E}}} = 0 \ \forall \ r\in \mathcal{R}^{\text{c}}$.
\end{lemma}

We end this section by pointing out that the terms composing the potentials have natural interpretations in terms of effective channels. 
The term $\mathbb{E}_z[\int dy\, \phi \log_2(\phi)]$ in $U_{\rm u}(E)$
is minus the conditional entropy $H(Y\vert Z)$ for the concatenation of the channels $\mathcal{N}(x\vert z\sqrt{1-E}, E)$
and $P_{\text{out}}(y|x)$ with a standardised input $\mathcal{N}(z\vert 0, 1)$. The term $S_{\rm u}( \Sigma( {E}))\log_2(B)$
is equal to minus the mutual information $I(\bf S;Y)$ for the Gaussian channel $\mathcal{N}(\mathbf{y}|\mathbf{s}, \tbf{I}\,\Sigma^2/\log_2(B))$ and input distribution $p_0(\bf s)$, up to a constant factor $-(2\ln(2))^{-1}$.

\section{Sketch of the proof of threshold saturation}\label{sec:proofsketch}
Monotonicity properties of the SE operators $T_{\rm u}$ and $T_{\rm c}$ are key elements in the analysis.  
\begin{definition}[Degradation]
A profile ${\tbf{E}}$ is degraded (resp. strictly degraded) w.r.t another one ${\tbf{G}}$, 
denoted as ${\tbf{E}} \succeq {\tbf{G}}$ (resp. ${\tbf{E}} \succ {\tbf{G}}$), if $E_r \ge  G_r \ \forall \ r$ 
(resp. if ${\tbf{E}} \succeq {\tbf{G}}$ and there exists some $r$ such that $E_r >  G_r$).
\label{def:degradation}
\end{definition}
%
%
%
\begin{lemma}
The SE operator of the coupled 
system maintains \emph{degradation in space}, i.e. if ${\tbf E} \succeq  {\tbf G}$, then $T_{\txt{c}}({\tbf E}) \succeq T_{\txt{c}}({\tbf{G}})$. It also maintains \emph{degradation in time}, i.e. $T_{\txt{c}}({\tbf E}^{(t)}) \preceq  {\tbf E}^{(t)} \Rightarrow T_{\txt{c}}({\tbf E}^{(t+1)}) \preceq  {\tbf E}^{(t+1)}$. Similarly $T_{\txt{c}}({\tbf E}^{(t)}) \succeq  {\tbf E}^{(t)}\Rightarrow T_{\txt{c}}({\tbf E}^{(t+1)}) \succeq  {\tbf E}^{(t+1)}$. Furthermore, the limiting profile ${\tbf E}^{(\infty)} \defeq T_{\txt{c}}^{(\infty)}({\tbf E}^{(0)})$
exists. These properties are verified by $T_{\txt{u}}$ for a scalar error as well.
\label{lemma:spaceDegrad}
\end{lemma}
\begin{IEEEproof}
Combining Lemma~\ref{lemma:SigmaIncreases} with \eqref{equ:sigmac} implies that if ${\tbf E} \succeq  {\tbf G}$, then
$\Sigma_c({\tbf E}) \geq \Sigma_c({\tbf G})\ \forall \ c$. The rest of the proof is similar to the one of Lemma~4.2 and Corollary~4.3 in \cite{barbierDiaMacris_isit2016}.
\end{IEEEproof}
\begin{figure}[!t]
\centering
\includegraphics[width=0.45\textwidth]{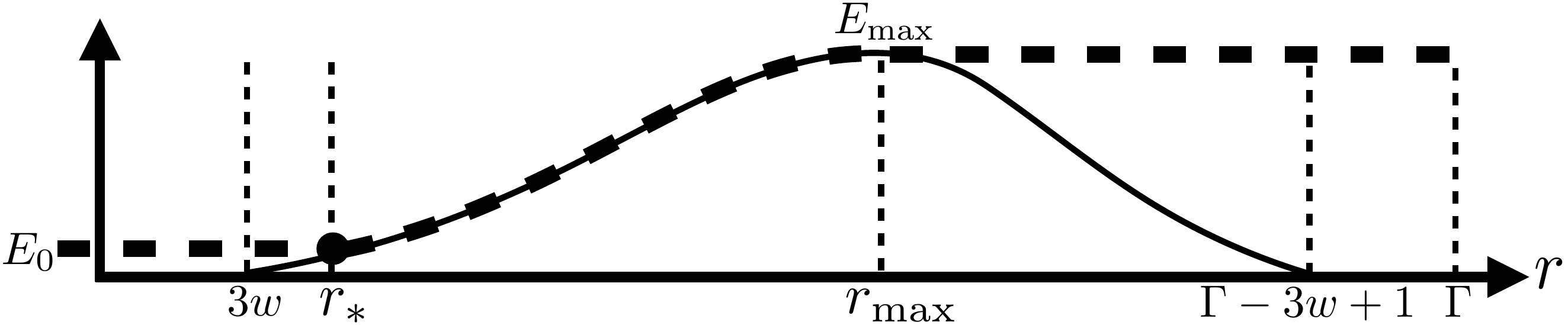}
\vspace*{-5pt}
\caption{A fixed point profile $\tbf{E}^{*}$ of the coupled SE (solid) is null $\forall \ r \le 3w$ and increases until $E_{{\rm max}}\in [0,1]$ at $r_{\rm max}\in \{3w+1:\Gamma-3w\}$. Then it decreases and is null $\forall \ r \ge \Gamma-3w+1$. The associated \emph{saturated} profile $\tbf E$ (dashed) starts at $E_0 \ \forall \ r \le r_*$, where $r_*$ is defined by: $E^{*}_{r}\le E_0 \ \forall \ r\le r_*$ and $E^{*}_{r'} > E_0 \ \forall \ r' > r_*$. Then it matches $\tbf E^{*} \ \forall \ r \in\{r_*:r_{\rm max}\}$ and saturates $E_{{\rm max}} \ \forall \ r \ge r_{\rm max}$. By construction ${\tbf E}$ is non decreasing and ${\tbf E} \succ {\tbf E}^{*}$.}
\label{fig:errorProfile}
\end{figure}

The pinning condition together with the monotonicity properties of the coupled SE imply that its {\it fixed point} profile ${\tbf E}^{*}$ must adopt a shape similar to Fig.~\ref{fig:errorProfile}. We associate to ${\tbf E}^{*}$ a {\it saturated profile} $\tbf E$ (see Fig.~\ref{fig:errorProfile}) that verifies by construction $\tbf E\succ{\tbf E}^{*}$. Thus $\tbf E$ serves as an upper bound in our proof.
\begin{definition}[Shift operator]
The \emph{shift operator} is defined componentwise as
$[\text{S}({\tbf E})]_1 \defeq E_0, \ [\text{S}({\tbf E})]_r \defeq  E_{r-1}$.
\end{definition}
\begin{lemma}
\label{lemma:quadFormBounded}
Let ${\tbf E}$ be a saturated profile. Then the coupled potential verifies $|F_{\text{c}}(\text{S}({\tbf E})) -F_{\text{c}}({\tbf E})| < K/w$, where $K$ is independent of $w$ and $\Gamma$.
\end{lemma}
\begin{IEEEproof}
The proof uses Lemmas~5.2, 5.3 and 5.4 of \cite{barbierDiaMacris_isit2016}, where Lemma~5.2 is implied by the present Lemma~\ref{lemma:fixedpointSE_extPot} and Lemma~5.3 remains valid as it depends only on the SC contruction. Lemma~5.4 can be shown to be true for any memoryless channel $P_{\text{out}}$ such that the function $g_{\text{out}} \defeq \partial_p \ln(\int dx P_{\text{out}}(y|x) \mathcal{N}(x|p,v))$ \cite{rangan2011generalized} is Lipschitz continuous in $p$ with Lipschitz constant independent of the coupling window.
\end{IEEEproof}
\begin{lemma}
Let ${\tbf E}$ be a saturated profile such that ${\tbf E}\succ {\tbf E}_0$. Then $F_{\text{c}}(\text{S}({\tbf E})) - F_{\text{c}}({\tbf E}) \leq -\Delta F_{\rm u}.$
\label{lemma:diffShited_directEval}
\end{lemma}
\begin{IEEEproof}
See the proof of Lemma~5.6 in \cite{barbierDiaMacris_isit2016}.
\end{IEEEproof}
\begin{thm}
Assume a spatially coupled SS code ensemble is used for communication through 
a memoryless channel. Fix $R<R_{\txt{pot}}$, $w>K/\Delta F_{\rm u}$ ($K$ is independent of $w$ and $\Gamma$)
and $\Gamma> 8w$ (such that the code is well defined). Then any fixed point profile ${\tbf{E}^*}$ of the coupled SE satisfies ${\tbf{E}}^* \prec {\tbf{E}}_0$.
\label{th:mainTheorem}
\end{thm}
\begin{IEEEproof}
It follows from Lemma~\ref{lemma:quadFormBounded} and \ref{lemma:diffShited_directEval} as in \cite{barbierDiaMacris_isit2016}.
\end{IEEEproof}
\begin{corollary}\label{cor:maincorollary}
By first 
taking $\Gamma \to \infty$ and then $w\to\infty$, the GAMP threshold of the coupled 
ensemble  satisfies $R_{\text c}\geq R_{{\rm pot}}$.
\end{corollary}

This result is a direct consequence of Theorem~\ref{th:mainTheorem} and Definition~\ref{def:AMPcoupled}. 
It says that 
the GAMP threshold of the coupled SS codes saturates to the potential threshold. 


We emphasize that Theorem~\ref{th:mainTheorem} and 
Corollary~\ref{cor:maincorollary} hold for a large class of estimation 
problems with random linear mixing \cite{rangan2011generalized}. Both the SE and potential formulations of Sec.~\ref{sec:stateandpot} as well as the proof
sketched in the present section are not restricted to SS codes. Indeed all the definitions and 
results are obtained for any memoryless channel $P_{\text{out}}$ and any factorizable (over $B$-d sections, $B\in\mathbb{N}$) prior over the message (or signal) $\bs$. 

%
\section{Large alphabet size analysis and \\connection with Shannon's capacity}
\label{sec:larg_B}
We now show that as the alphabet size $B$ increases, the potential threshold of SS codes approaches Shannon's capacity $R_{\rm pot}^{\infty}\defeq\lim_{B\to \infty} R_{\rm pot} = C$, and also that $\lim_{B\to \infty}E_0 = 0$. Note that these are static properties of the code independent of the decoder. But note also that the threshold 
saturation established in Corollary~\ref{cor:maincorollary} for SC-SS codes implies that optimal decoding can actually be performed using the GAMP decoder, i.e. $\lim_{B\to \infty} R_{\text c} = C$, since $R_{\text c} \leq C$.

%

\begin{figure}[t!]
\centering
\includegraphics[width=0.241\textwidth, height=110pt, trim={0pt 3 3 0},clip]{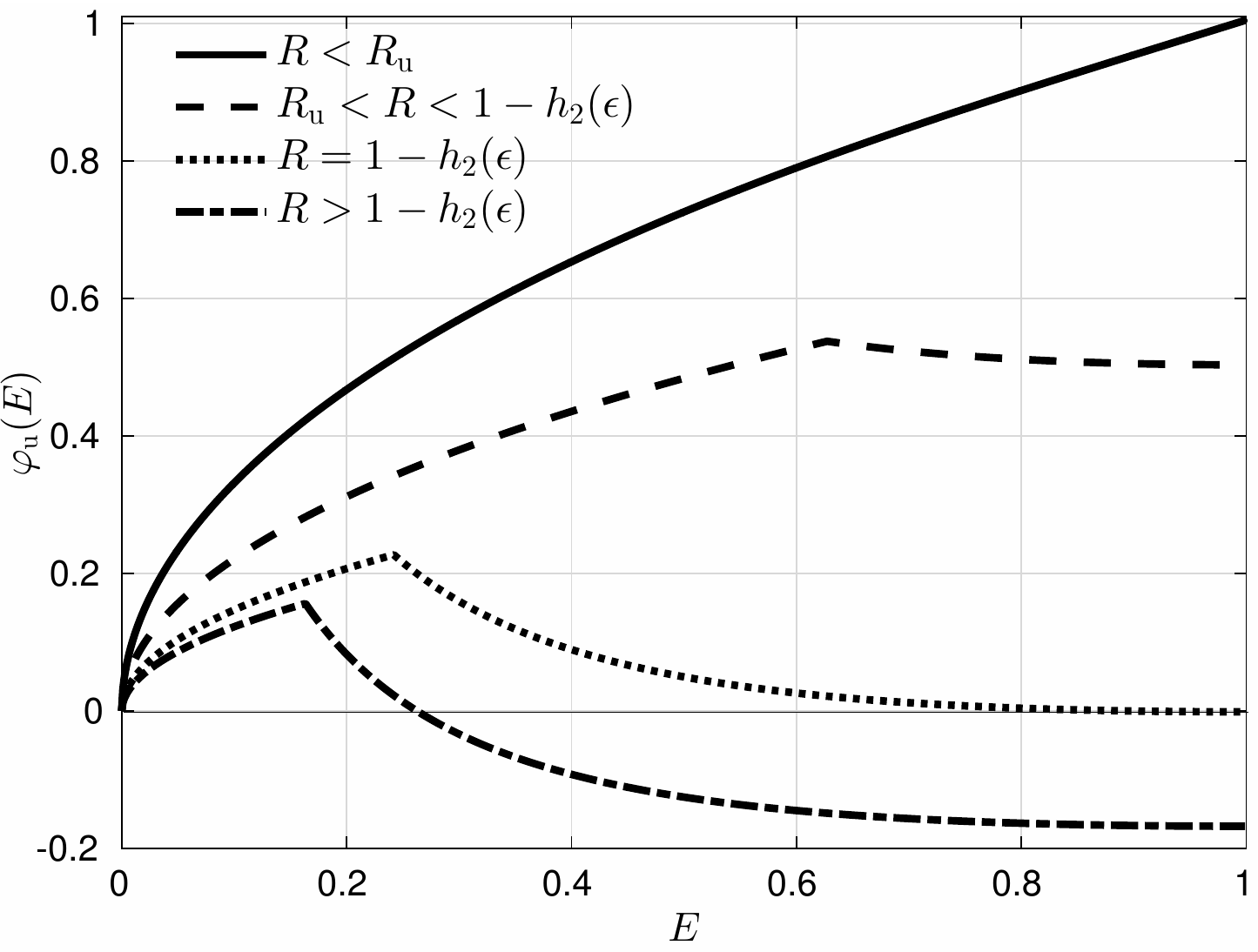}
\centering
\includegraphics[width=0.241\textwidth, height=108.7pt, trim={0pt 3 3 0},clip]{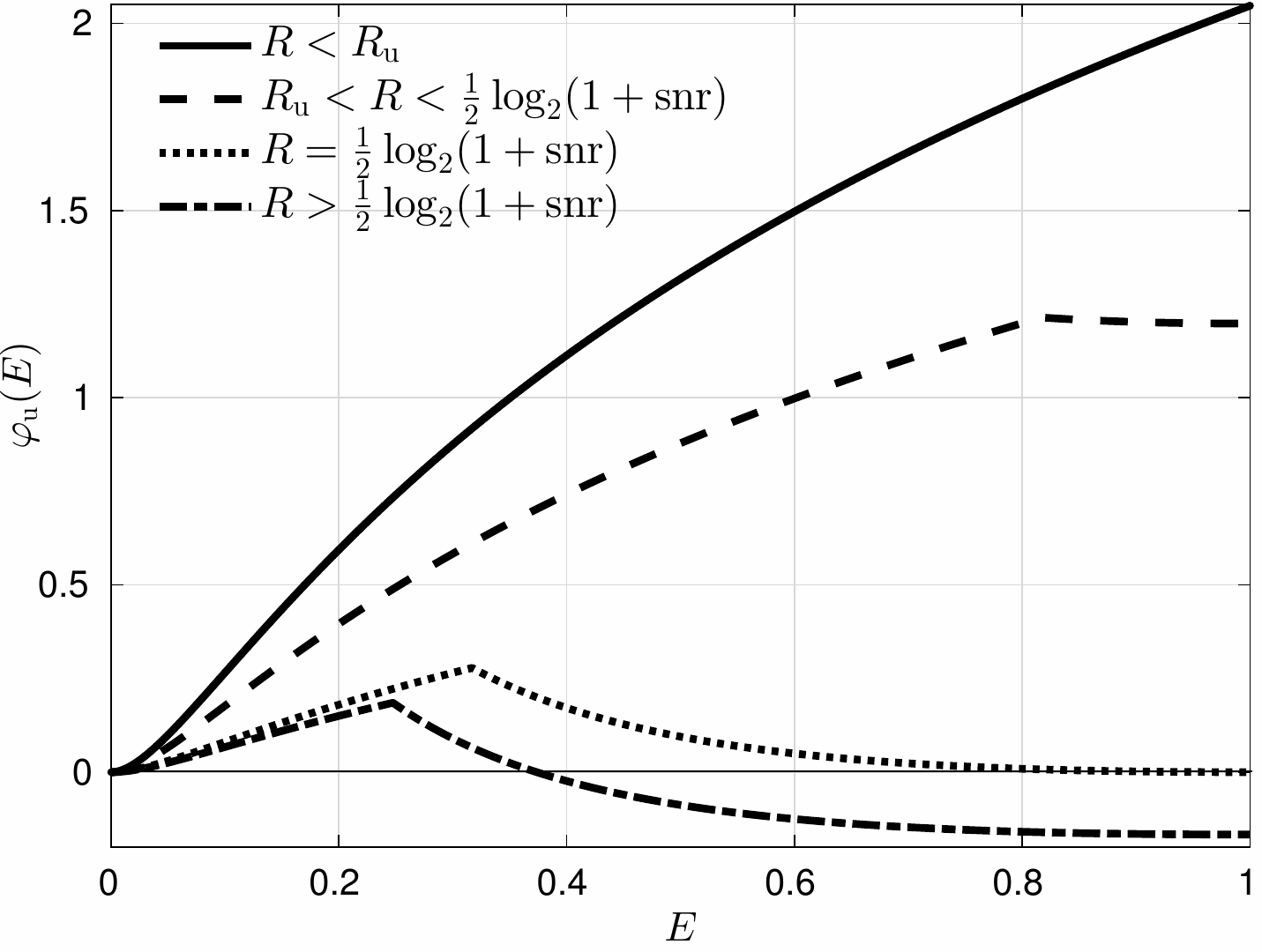}
\vspace*{-16pt}
\caption{The large alphabet potential $\varphi_{\rm u}(E)$ (\ref{eq:largeBpot}) as a function of the MSE for the BSC (left) and AWGN (right) channels with $\epsilon=0.1$ and ${\rm snr}=10$ respectively. $\varphi_{\rm u}(E)$ is scaled such that $\varphi_{\rm u}(0)=0$. For $R$ below the GAMP threshold $R_{\rm u}$, there is a unique minimum at $E=0$ while just above $R_{\rm u}$, this minimum coexists with a local one at $E=1$. At the optimal threshold of the code, that coincides with the Shannon capacity, the two minima are equal. Then, for $R>C$ the minimum at $E=1$ becomes the global one, and thus decoding is impossible.}
\label{fig:largeB}
\end{figure}

Lemma \ref{lemma:fixedpointSE_extPot} implies that the underlying system's potential contains all the information about $R_{\rm pot}$ and $R_{\rm u}$. Hence, we proceed by computing $\varphi_{\rm u}(E) \defeq \lim_{B\to\infty} F_{\rm u}(E)$  \cite{phdBarbier},
\begin{align} \label{eq:largeBpot}
\varphi_{\rm u}(E) = U_{\rm u}(E) - {\rm max}\Big(0,1 - \frac{1}{2\ln(2)\Sigma(E)^2}\Big). 
\end{align}
The analysis of $\varphi_{\rm u}(E)$ for $E\in[0,1]$ shows that the only possible minima are at $E=0$ and $E=1$, which implies that the error floor $E_0$ vanishes as $B$ increases (Fig. \ref{fig:largeB}). One can show that if $\Sigma(E)^2 < (2\ln(2))^{-1}\ \forall\ E\in[0,1]$, which corresponds to the region $R<(2\ln(2))^{-1}\mathbb{E}_{p|1} [\mathcal{F}(p|1)]$ for any fixed memoryless channel, then $\varphi_{\rm u}(E)$ has a unique minimum at $E = 0$. Similarly for $R>(2\ln(2))^{-1}\mathbb{E}_{p|0} [\mathcal{F}(p|0)]$ there is a unique minimum at $E=1$. In the \emph{intermediate region} both minima coexist. Therefore, we identify 
\begin{align}
R_{\rm u}^{\infty}\defeq \lim_{B\to \infty} R_{\rm u} = \frac{\mathbb{E}_{p|1} [\mathcal{F}(p|1)]}{2\ln(2)} = \frac{\mathcal{F}(0|1)}{2\ln(2)}.
\end{align}
Since $R_{\rm pot}$ is defined by the point where $\Delta F_{\rm u}$ switches sign (Definition \ref{def:potThresh}), $R_{\rm pot}^{\infty}$ can be obtained by equating the two minima of $\varphi_{\rm u}(E)$. Setting $\varphi_{\rm u}(1) = \varphi_{\rm u}(0)$ yields
\begin{align}\label{eq:large_B}
R_{\rm pot}^{\infty}= &- \int dy \mathcal{D}z P_{\text{out}}(y|z) \log_2\Big( \int \mathcal{D}\tilde z P_{\text{out}}(y|\tilde z)\Big) \nonumber \\
&+ \int dy \mathcal{D}z P_{\text{out}}(y|z) \log_2\Big(P_{\text{out}}(y|z)\Big),
\end{align}
where $\mathcal{D}z$ is a standard Gaussian distribution. We will now recognize that this expression is the Shannon capacity of $W$ for a proper choice of the map $\pi$.

Let $\mathcal{A}$ and $\mathcal{B}$ be the input and output alphabet of $W$ respectively, where $\mathcal{A}, \mathcal{B} \subseteq \mathbb{R}$ are defined over discrete or continuous supports. Call $\mathcal{P}$ the capacity-achieving input distribution associated with $W$.
Choose $\pi:\mathbb{R}\to\mathcal{A}$ such that $i)$ $P_{\text{out}}(y|z) = W(y|\pi(z))$ and $ii)$ if $z\sim\mathcal{N}(z|0,1)$, then $\pi(z) \sim \mathcal{P}$. This map converts a standard Gaussian random variable $z$ onto a channel-input random variable $\pi(z)=a$ with capacity-achieving distribution $\mathcal{P}(a)$. Note that $\pi$ can be viewed equivalently as part of the code or of the channel. 

Now using the relation $\int \mathcal{D}z P_{\text{out}}(y|z) = \int \mathcal{D}z W(y|\pi(z)) =  \int da \mathcal{P}(a) W(y|a)$, (\ref{eq:large_B}) can be expressed equivalently as
\begin{align}\label{eq:large_B_symm}
R_{\rm pot}^{\infty}= &- \int dy da \mathcal{P}(a) W(y|a) \log_2\Big( \int d\tilde a \mathcal{P}(\tilde a) W(y|\tilde a)\Big) \nonumber \\
&+ \int dy da \mathcal{P}(a) W(y|a) \log_2\Big(W(y|a)\Big).
\end{align}
The first term in (\ref{eq:large_B_symm}) is nothing but the Shannon entropy $H(Y)$ of the channel output-distribution, while the second term is the negative of the conditional entropy $H(Y|A)$ of the channel-output distribution given the input $A=\pi(Z)$, that has capacity-achieving distribution. Thus, $R_{\rm pot}^{\infty}$ is the Shannon capacity of $W$.
Combining this result with Corollary~\ref{cor:maincorollary}, we can assert that SC-SS codes allow to communicate reliably up to Shannon's capacity over any memoryless channel under low complexity GAMP decoding.

\begin{figure}[b!]
\centering
\includegraphics[width=0.24\textwidth, height=110pt, trim={0pt -1.3 3 0},clip]{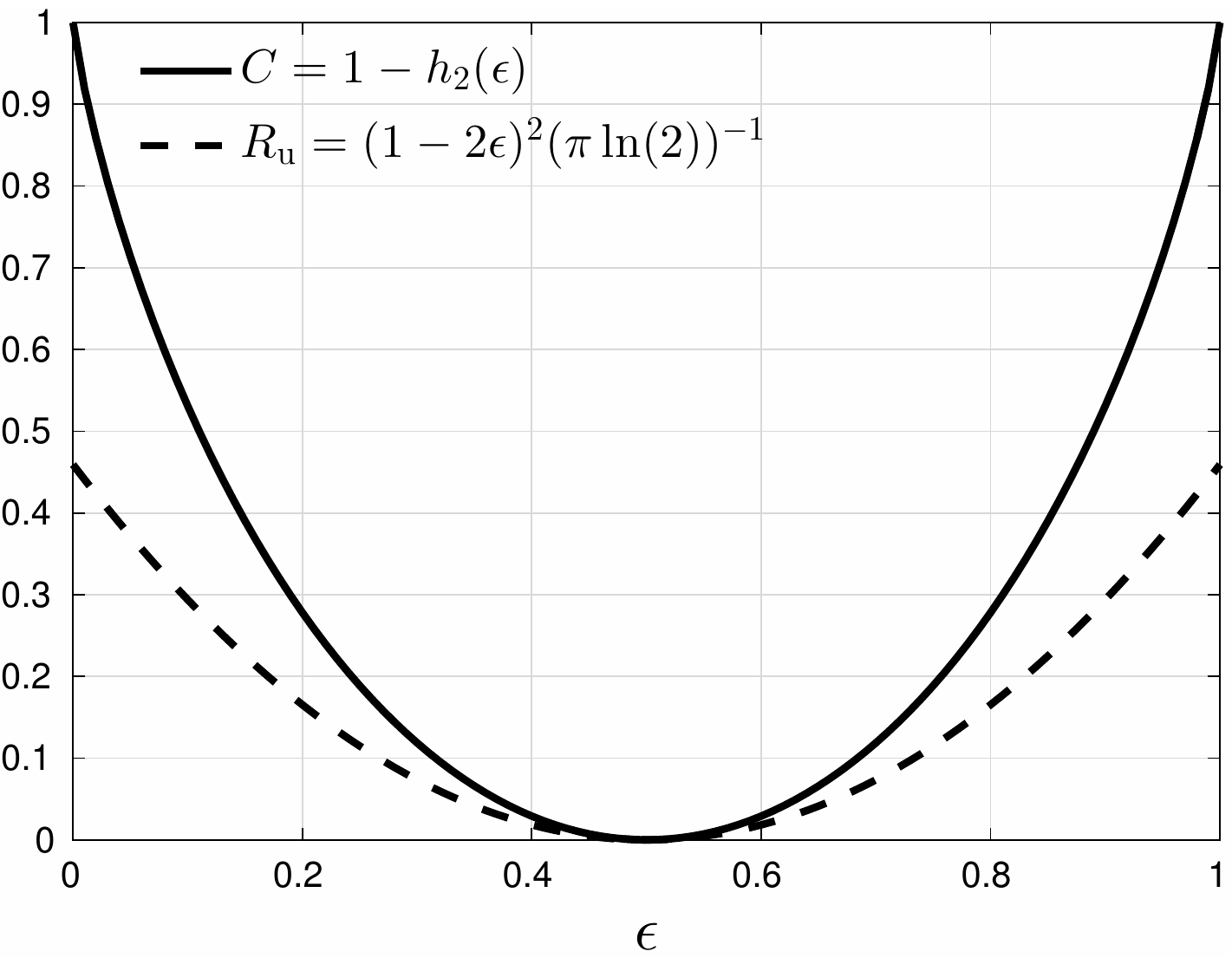}
\includegraphics[width=0.24\textwidth, height=110pt, trim={0pt 0 1 0},clip]{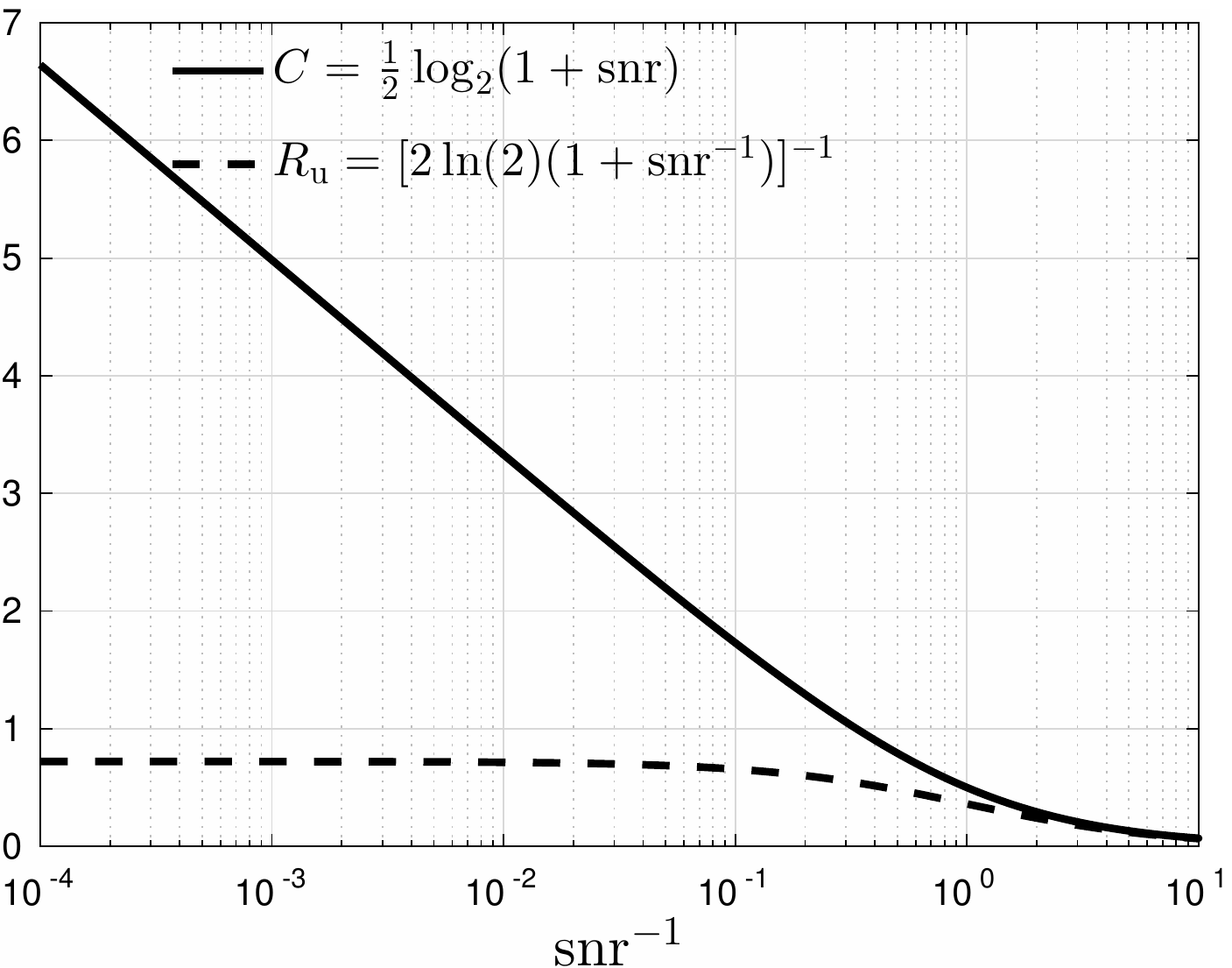}
\vspace*{-16pt}
\caption{Large alphabet limits of the capacities and GAMP thresholds for the BSC (left) and AWGN (right) channels.}
\label{fig:capacities}
\end{figure}

\begin{figure}[t]
\centering
\includegraphics[width=0.28\textwidth, height=110pt, trim={0pt 2 3 0},clip]{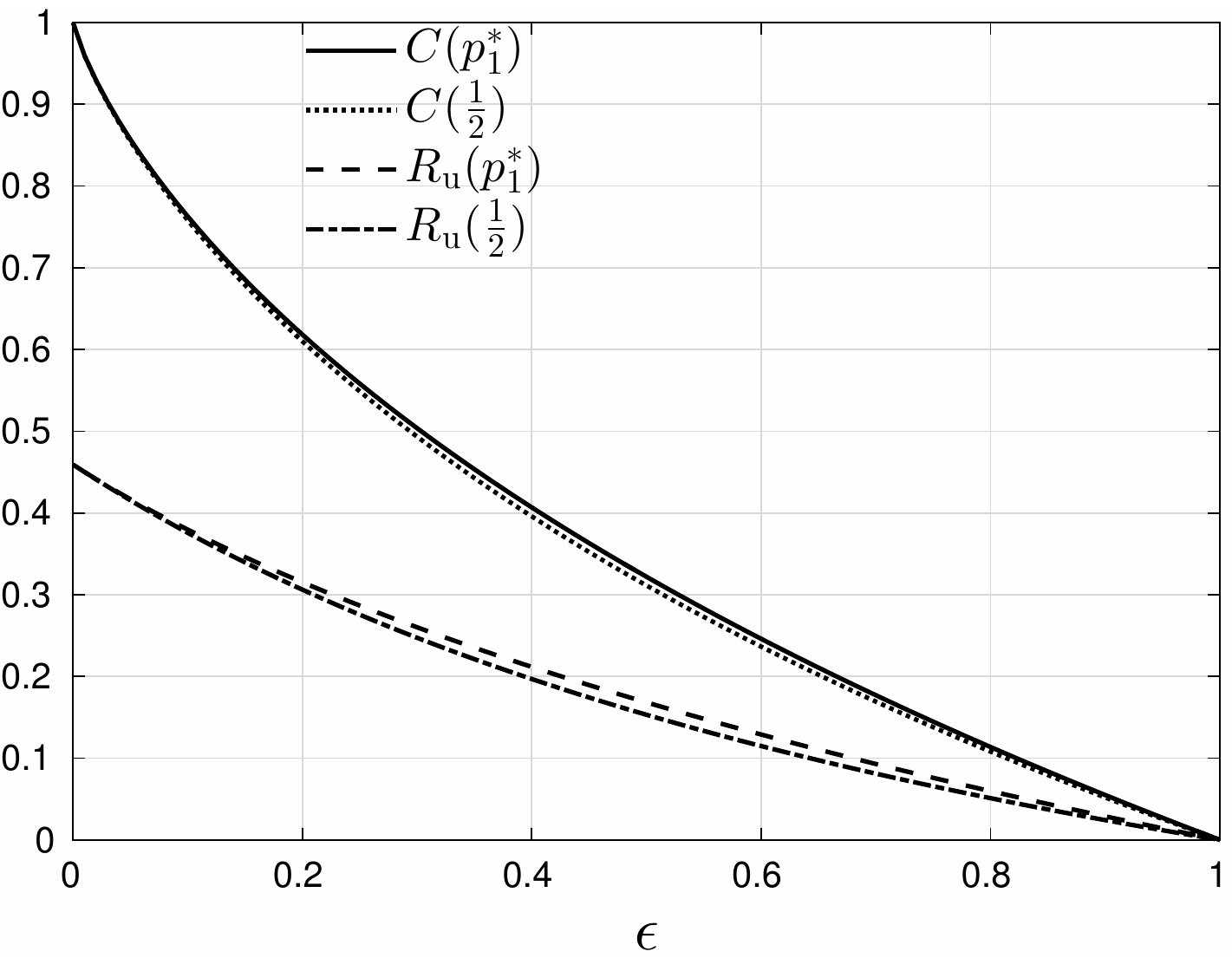}
\vspace*{-5pt}
\caption{Capacity and GAMP threshold of the Z channel. $C(p_1^{*})$ and $R_{\rm u}(p_1^{*})$ are the values under capacity-achieving input distribution, whereas $C(\frac{1}{2})$ and $R_{\rm u}(\frac{1}{2})$ are the values under uniform distribution.}
\label{fig:capacity_Z}
\end{figure}

But how to find the proper map $\pi$ for a given memoryless channel? In the case of discrete input memoryless symmetric channels, Shannon's capacity can be attained by inducing a uniform input distribution $\mathcal{P} = \mathcal{U}_\mathcal{A}$. Let us call $q$ the cardinality of $\mathcal{A}=\{a_1:a_q\}$. In this case the mapping $\pi$ is simply $\pi(z) = a_i$ if $z\in \, ]z_{(i-1)/q}, z_{i/q}]$, where $z_{i/q}$ is the $i^{th}$ $q$-quantile of the Gaussian distribution, with $z_{0} = -\infty, z_{1}=\infty$. For asymmetric channels, one can use some standard methods such as Gallager's mapping or more advanced ones \cite{MondelliUrbankeHassani_assymetricChannels} that introduce bias in the channel-input distribution in order to match the capacity-achieving one.
%
%
We now illustrate these findings, depicted for various channels in Fig.~\ref{fig:capacities} and Fig.~\ref{fig:capacity_Z}.  

\emph{\textbf{AWGN channel}}: We start showing that our results for the AWGN channel \cite{barbierDiaMacris_isit2016} are a special case of the present general framework. No map $\pi$ is required and the Shannon capacity is directly obtained from (\ref{eq:large_B}) because the capacity-achieving input distribution for the AWGN channel is Gaussian. Thus, by plugging $P_{\text{out}}(y|z)=\mathcal{N}(y|z,1/{\rm snr})$ in (\ref{eq:large_B}), one recovers the Shannon capacity $R_{\rm pot}^{\infty} = \frac{1}{2}\log_2(1+{\rm snr})$. Furthermore, one obtains $R_{\rm u}^{\infty}=[2\ln(2)(1+1/\rm {snr})]^{-1}$.

\emph{\textbf{BSC channel}}: The binary symmetric channel (BSC) with flip probability $\epsilon$ has transition probability $W(y|a) = (1-\epsilon) \delta(y-a) + \epsilon \delta(y+a)$, where both $y,a \in \{-1,1\}$. The proper map is $\pi(z) = {\rm sign}(z)$ since it induces uniform input distribution $\mathcal{U}_{\mathcal{A}}=1/2$. So by plugging $W$ and $\mathcal{U}_{\mathcal{A}}$ in (\ref{eq:large_B_symm}), or equivalently $P_{\text{out}}(y|z) = (1-\epsilon) \delta(y-\pi(z)) + \epsilon \delta(y+\pi(z))$ into (\ref{eq:large_B}), one obtains the Shannon capacity of the BSC channel $R_{\rm pot}^{\infty} = 1-h_2(\epsilon)$ where $h_2$ is the binary entropy function. This map also gives $R_{\rm u}^{\infty}=(\pi\ln(2))^{-1}(1-2\epsilon)^{2}$.

\emph{\textbf{BEC channel}}: Note that the binary erasure channel (BEC) is also symmetric. Therefore, the same mapping $\pi(z) = {\rm sign}(z)$ is used and leads to the Shannon capacity $R_{\rm pot}^{\infty} =1-\epsilon$, where $\epsilon$ is the erasure probability, and $R_{\rm u}^{\infty}=(\pi\ln(2))^{-1}(1-\epsilon)$.

\emph{\textbf{Z channel}}: The Z channel is the extremal discrete asymmetric channel, in the sense that it represents the ``worst'' one. It has binary input and output $\in\{-1,1\}$ with transition probability $W(y|a) = \delta(a-1)\delta(y-a) + \delta(a+1)[(1-\epsilon) \delta(y-a) + \epsilon \delta(y+a)]$, where $\epsilon$ is the flip probability of the $-1$ input. The map $\pi(z)={\rm sign}(z)$ leads to the \emph{symmetric capacity} of the Z channel $R_{\rm pot}^{\infty} =h_2((1-\epsilon)/2) - h(\epsilon)/2$, that is the input-output mutual information when the input is uniformly distributed, and $R_{\rm u}^{\infty}=[\pi\ln(2)(1+\epsilon)]^{-1}(1-\epsilon)$. This expression differs from Shannon's capacity. However, one can introduce bias in the input distribution and hence match the capacity-achieving one. To do so, the proper map defined in terms of the $Q$-function is $\pi(z) = {\rm sign}(z - Q^{-1}(p_1))$, where $p_1$ is the input probability of the bit $1$. By optimizing over $p_1$, one can obtain the Shannon's capacity of the Z channel for $p_1^{*} = 1 - [(1-\epsilon)(1+2^{h_2(\epsilon)/(1-\epsilon)})]^{-1}$.
\section{Open challenges} \label{sec:openChallenges}
We end up pointing some open problems. In order to have a fully rigorous capacity achieving scheme over any memoryless channel, using SC-SS codes and GAMP decoding, it must be shown that the SE tracks the asymptotic performance of GAMP. We conjecture that it is indeed the case and that the proof follows from the method of \cite{BayatiMontanari10}, then extended in \cite{rush2015capacity} for power allocated SS codes. It is also desirable to consider practical coding schemes, using Hadamard-based operators or more generally, row-othogonal matrices. Another important point is to estimate at what rate the error floor vanishes when $B$ increases. Finally, the finite size effects should be considered in order to assess the real potential of these codes. We plan to settle these questions in future works.
\section*{Acknowledgments}
J.B and M.D acknowledge funding from the Swiss National Science Foundation grant num. 200021-156672. We thank Florent Krzakala, Rüdiger Urbanke and Christophe Schülke for helpful discussions.
\ifCLASSOPTIONcaptionsoff
\fi
\bibliographystyle{IEEEtran}
\bibliography{IEEEabrv,bibliography}

\begin{thebibliography}{10}
\providecommand{\url}[1]{#1}
\csname url@samestyle\endcsname
\providecommand{\newblock}{\relax}
\providecommand{\bibinfo}[2]{#2}
\providecommand{\BIBentrySTDinterwordspacing}{\spaceskip=0pt\relax}
\providecommand{\BIBentryALTinterwordstretchfactor}{4}
\providecommand{\BIBentryALTinterwordspacing}{\spaceskip=\fontdimen2\font plus
\BIBentryALTinterwordstretchfactor\fontdimen3\font minus
  \fontdimen4\font\relax}
\providecommand{\BIBforeignlanguage}[2]{{%
\expandafter\ifx\csname l@#1\endcsname\relax
\typeout{** WARNING: IEEEtran.bst: No hyphenation pattern has been}%
\typeout{** loaded for the language `#1'. Using the pattern for}%
\typeout{** the default language instead.}%
\else
\language=\csname l@#1\endcsname
\fi
#2}}
\providecommand{\BIBdecl}{\relax}
\BIBdecl

\bibitem{barbierDiaMacris_isit2016}
\BIBentryALTinterwordspacing
J.~{Barbier}, M.~{Dia}, and N.~{Macris}, ``{Proof of Threshold Saturation for
  Spatially Coupled Sparse Superposition Codes},'' \emph{ArXiv e-prints}, Mar.
  2016. [Online]. Available: \url{http://arxiv.org/pdf/1603.01817v1.pdf}
\BIBentrySTDinterwordspacing

\bibitem{barron2010sparse}
A.~Barron and A.~Joseph, ``Toward fast reliable communication at rates near
  capacity with gaussian noise,'' in \emph{Information Theory Proceedings
  (ISIT), 2010 IEEE International Symposium on}, June 2010, pp. 315--319.

\bibitem{JosephB14}
A.~Joseph and A.~R. Barron, ``Fast sparse superposition codes have near
  exponential error probability for {R}<{C},'' \emph{{IEEE} Trans. on
  Information Theory}, vol.~60, no.~2, pp. 919--942, 2014.

\bibitem{barron2012high}
A.~R. Barron and S.~Cho, ``High-rate sparse superposition codes with
  iteratively optimal estimates,'' in \emph{Information Theory Proceedings
  (ISIT), 2012 IEEE International Symposium on}.\hskip 1em plus 0.5em minus
  0.4em\relax IEEE, 2012, pp. 120--124.

\bibitem{barbier2014replica}
J.~Barbier and F.~Krzakala, ``Replica analysis and approximate message passing
  decoder for superposition codes,'' in \emph{Information Theory Proceedings
  (ISIT), 2014 IEEE International Symposium on}, 2014.

\bibitem{barbierSchulkeKrzakala}
J.~Barbier, C.~Schülke, and F.~Krzakala, ``Approximate message-passing with
  spatially coupled structured operators, with applications to compressed
  sensing and sparse superposition codes,'' \emph{Journal of Statistical
  Mechanics: Theory and Experiment}, vol. 2015, no.~5, 2015.

\bibitem{BarbierK15}
\BIBentryALTinterwordspacing
J.~Barbier and F.~Krzakala, ``Approximate message-passing decoder and
  capacity-achieving sparse superposition codes,'' 2015. [Online]. Available:
  \url{http://arxiv.org/abs/1503.08040}
\BIBentrySTDinterwordspacing

\bibitem{KrzakalaMezard12}
F.~Krzakala, M.~M\'ezard, F.~Sausset, Y.~Sun, and L.~Zdeborov\'a,
  ``Probabilistic reconstruction in compressed sensing: Algorithms, phase
  diagrams, and threshold achieving matrices,'' \emph{Journal of Statistical
  Mechanics: Theory and Experiment}, vol. P08009, 2012.

\bibitem{CaltagironeZ14}
F.~Caltagirone and L.~Zdeborov{\'{a}}, ``Properties of spatial coupling in
  compressed sensing,'' \emph{CoRR}, vol. abs/1401.6380, 2014.

\bibitem{rush2015capacity}
C.~Rush, A.~Greig, and R.~Venkataramanan, ``Capacity-achieving sparse
  regression codes via approximate message passing decoding,'' in
  \emph{Information Theory (ISIT), 2015 IEEE International Symposium on}, June
  2015, pp. 2016--2020.

\bibitem{YedlaJian12}
A.~Yedla, Y.-Y. Jian, P.~S. Nguyen, and H.~D. Pfister, ``A simple proof of
  threshold saturation for coupled scalar recursions,'' in \emph{7th
  International Symposium on Turbo Codes and Iterative Information Processing
  (ISTC)}, 2012, pp. 51--55.

\bibitem{PfisterMacrisBMS}
S.~Kumar, A.~J. Young, N.~Macris, and H.~D. Pfister, ``Threshold saturation for
  spatially-coupled ldpc and ldgm codes on bms channels,'' \emph{IEEE Trans. on
  Information Theory}, vol.~60, pp. 7389--7415, 2013.

\bibitem{6887298}
A.~Yedla, Y.-Y. Jian, P.~Nguyen, and H.~Pfister, ``A simple proof of maxwell
  saturation for coupled scalar recursions,'' \emph{Information Theory, IEEE
  Trans. on}, vol.~60, no.~11, pp. 6943--6965, 2014.

\bibitem{rangan2011generalized}
S.~Rangan, ``Generalized approximate message passing for estimation with random
  linear mixing,'' in \emph{Information Theory Proceedings (ISIT), 2011 IEEE
  International Symposium on}.\hskip 1em plus 0.5em minus 0.4em\relax IEEE,
  2011, pp. 2168--2172.

\bibitem{fisherInfoProperties}
\BIBentryALTinterwordspacing
P.~Zegers, ``Fisher information properties,'' \emph{Entropy}, vol.~17, no.~7,
  p. 4918, 2015. [Online]. Available: \url{http://mdpi.com/1099-4300/17/7/4918}
\BIBentrySTDinterwordspacing

\bibitem{phdBarbier}
\BIBentryALTinterwordspacing
J.~Barbier, ``Statistical physics and approximate message-passing algorithms
  for sparse linear estimation problems in signal processing and coding
  theory,'' Ph.D. dissertation, Université Paris Diderot, 2015. [Online].
  Available: \url{http://arxiv.org/abs/1511.01650}
\BIBentrySTDinterwordspacing

\bibitem{MondelliUrbankeHassani_assymetricChannels}
M.~Mondelli, R.~Urbanke, and S.~H. Hassani, ``How to achieve the capacity of
  asymmetric channels,'' in \emph{Communication, Control, and Computing, 2014
  Allerton Conference on}.\hskip 1em plus 0.5em minus 0.4em\relax IEEE, 2014,
  pp. 789--796.

\bibitem{BayatiMontanari10}
M.~Bayati and A.~Montanari, ``The dynamics of message passing on dense graphs,
  with applications to compressed sensing,'' \emph{IEEE Trans. on Information
  Theory}, vol.~57, no.~2, pp. 764 --785, 2011.

\end{thebibliography}
\end{document}